\documentclass[aps,  
prc,                 
showpacs,            
twocolumn,           
preprintnumbers,
nofootinbib,
floatfix,            
superscriptaddress]  
{revtex4-1}

\usepackage{graphicx} 
\usepackage{epsfig}   
\usepackage{bm}       
\usepackage{amsmath}
\usepackage{slashed}  
\usepackage{hyperref}
\usepackage{physics}  
\usepackage[usenames,dvipsnames,svgnames,table]{xcolor}
\usepackage[utf8]{inputenc}

\hypersetup{colorlinks,linkcolor=MidnightBlue,citecolor=MidnightBlue,urlcolor=MidnightBlue}
\hypersetup{breaklinks=true}

\begin{document}


\title{Influence of electromagnetic fields in proton-nucleus collisions at relativistic energy}


\author{Lucia Oliva}
\affiliation{GSI Helmholtzzentrum f\"{u}r Schwerionenforschung GmbH, Planckstr. 1, 64291 Darmstadt, Germany}
\affiliation{Institut f\"{u}r Theoretische Physik, Johann Wolfgang Goethe-Universit\"{a}t, Max-von-Laue-Str. 1, 60438 Frankfurt am Main, Germany}

\author{Pierre Moreau}
\affiliation{Institut f\"{u}r Theoretische Physik, Johann Wolfgang Goethe-Universit\"{a}t, Max-von-Laue-Str. 1, 60438 Frankfurt am Main, Germany}
\affiliation{Department of Physics, Duke University, Durham, North Carolina 27708, USA}

\author{Vadim Voronyuk}
\affiliation{Joint Institute for Nuclear Research, Joliot-Curie 6, 141980 Dubna, Moscow region, Russia}
\affiliation{Bogolyubov Institute for Theoretical Physics, Metrolohichna str. 14-b, 03143 Kiev, Ukraine}

\author{Elena Bratkovskaya}
\affiliation{GSI Helmholtzzentrum f\"{u}r Schwerionenforschung GmbH, Planckstr. 1, 64291 Darmstadt, Germany}
\affiliation{Institut f\"{u}r Theoretische Physik, Johann Wolfgang Goethe-Universit\"{a}t, Max-von-Laue-Str. 1, 60438 Frankfurt am Main, Germany}


\begin{abstract}
We study proton-gold collisions at RHIC energy $\sqrt{s_{NN}}=200$ GeV within the Parton-Hadron-String Dynamics (PHSD) off-shell transport approach, investigating the influence of the intense electromagnetic fields generated in these small systems.
We show the space-time evolution of the magnetic and electric components, 
emphasizing the huge values of the latter one, in particular the electric field $E_x$ along the impact parameter direction whose magnitude is comparable to the magnetic field $B_y$ perpendicular to the reaction plane.
We find a fair agreement of the charged particle pseudorapidity density of the high-multiplicity events with respect to the experimental result of the PHENIX Collaboration. Focusing on the most central collision, we show rapidity distributions and spectra as well as the flow coefficients $v_1$ and $v_2$ and we discuss the impact of the electromagnetic fields on identified particle observables. We compute the directed flow $v_1$ of $\pi^+$, $\pi^-$, $K^+$, $K^-$ for collisions at fixed impact parameter and predict that an electromagnetically-induced splitting in the $v_1$ of positively and negatively charged particles is generated in the Au-going side of p+Au reaction mainly driven by the huge $E_x$ component. We find that this effect is stronger for the strange mesons and increases for increasing impact parameter. Furthermore, we highlight the amount of directed flow generated in the deconfined phase, finding that it constitutes the main contribution in the central rapidity region, especially for kaons. Thus, we support the idea that the directed flow is a promising probe for the electromagnetic fields generated in relativistic nuclear collisions and show that in proton-induced reactions the electric component along the impact parameter axis is the primary origin of a charge-odd $v_1$ of pions and kaons.

\end{abstract}


\maketitle

\section{Introduction}
\label{intro}

One of the most surprising discoveries of heavy ion experiments at high energy was that the Quark-Gluon Plasma (QGP) created after the collision is a strongly-coupled system that exhibits a fluid behaviour with the development of anisotropic collective flows. Indeed, experiments have observed a large value of the elliptic flow $v_2$, which is a measure of the asymmetry in transverse momentum space and is characterized by the second-order harmonic of the Fourier expansion of the particle azimuthal distribution with respect to the reaction plane \cite{Poskanzer:1998yz}.
The generation of an elliptic flow is connected to the transport properties of the matter created after the collision, such as the shear viscosity over entropy density ratio $\eta/s$; the estimated value is very close to the lower bound $\eta/s=1/4\pi$ conjectured for a strongly interacting system \cite{Kovtun:2004de} establishing the nature of QGP as a nearly perfect fluid.

The QGP was initially expected to be formed only in relativistic collisions of two heavy ions and small colliding systems, such as proton-proton and proton-nucleus collisions, were only regarded as control measurements.
However, in recent years, examining high-multiplicity proton-nucleus and deuteron-nucleus collisions at LHC and RHIC energies respectively it turned out that also in these small systems particles had a clear preference to be emitted along common transverse directions, though on a smaller scale respect to nucleus-nucleus collisions \cite{CMS:2012qk,Aad:2012gla,Abelev:2012ola,Adare:2013piz}.
Thus a scientific debate started on the nature of this azimuthal asymmetry, whether it is related to the formation of QGP droplets or due to early-time momentum correlations. The recent experimental and theoretical results add to a growing body of evidence that even in small systems the QGP is created and during its expansion it translates efficiently the initial-state geometric eccentricity into a final-state momentum anisotropy \cite{PHENIX:2018lia}; see e.g. Ref.~\cite{Nagle:2018nvi} for a review.

In relativistic collisions of two heavy ion the $v_2$ shows a strong dependence with the impact parameter since it is mainly related to the global almond shape of the overlap region. However, it acquires a contribution also from the initial fluctuations of nucleon positions in the overlap region. These geometrical fluctuations give also rise to the odd harmonics $v_1$, $v_3$, $v_5$, etc... of the Fourier expansion of the particle azimuthal distribution. Nevertheless, the directed flow $v_1$, which refers to a collective sidewards deflection of particles, is also strongly connected to the generation of vortical patterns in the QGP due to the angular momentum of the system and to the electromagnetic fields produced in the initial stage of the collision.

Indeed, in the last decade it has been established that extremely intense electromagnetic fields are produced in non-central relativistic heavy-ion collisions mainly due to the motion of spectator charges \citep{Skokov:2009qp, Voronyuk:2011jd}.
In particular, the magnetic field in the very early stage of collisions at top RHIC and LHC energies can reach values of $|eB_y|\sim 5-50\,m_{\pi}^2$, which correspond to about $10^{18}-10^{19}$ Gauss, i.e. some order of magnitude higher than that expected to be produced in magnetars.
The directed flow of both light \cite{Voronyuk:2014rna, Toneev:2016bri, Gursoy:2014aka, Gursoy:2018yai} and heavy mesons \cite{Das:2016cwd, Coci:2019nyr, Chatterjee:2018lsx} has been considered as a promising probe to characterize the generated electric and magnetic fields and the recent experimental results from STAR \cite{Adamczyk:2017nxg, Adam:2019wnk} and ALICE \cite{Acharya:2019ijj} Collaborations challenge the theoretical understanding of the $v_1$ and its connection to the electromagnetic fields in different collision systems and energies.

The electromagnetic fields have never been studied in the context of small colliding systems on a basis of microscopic transport approaches.
With the present work we aim at filling this gap, focusing on proton-gold collisions at RHIC energy of $\sqrt{s_{NN}}=200$ GeV.
Our main goal is the study of the fields produced in this strongly asymmetric system, investigating their effect on the collective behaviour of the created matter, quantifying in particular the electromagnetically-induced splitting in the charge-dependent directed flow of the most abundant hadron species at top RHIC energy.
\\
To this end we perform simulations with the Parton-Hadron-String Dynamics (PHSD), which is a microscopic off-shell transport approach that describes the full space-time evolution of a relativistic nuclear collision from the initial hard scatterings and string formation through the dynamical onset of the  deconfined QGP phase to the hadronization and subsequent interactions in the hadronic phase \cite{Cassing:2008sv, Cassing:2009vt, Bratkovskaya:2011wp}.
PHSD includes also the dynamical formation and evolution of the electromagnetic fields during the collision and their influence on quasiparticle propagation as well as the back-reaction of particle dynamics on the fields \citep{Voronyuk:2011jd, Voronyuk:2014rna, Toneev:2016bri}.
Besides an analysis of the properties of p+Au collisions and its final particle distributions, we present our predictions for the $v_1$ of charged and identified hadrons and discuss the influence of electromagnetic fields, which lead to a separation of positively and negatively charged mesons along the impact parameter axis that is far more pronounced in the Au-going side.

In an early work, PHSD has been applied for studying the dynamics of p+A reaction at ultra-relativistic energies \cite{Konchakovski:2014wqa}. The collective anisotropies in these small systems have been investigated in many other theoretical works \cite{Bozek:2011if, Bozek:2013uha, Werner:2013ipa, Kozlov:2014fqa, Shen:2016zpp, Weller:2017tsr, Bzdak:2014dia, Greif:2017bnr, Sun:2019gxg}.
However, the influence of the electromagnetic fields on particle dynamics and the first flow harmonic $v_1$ have not been explored in those studies.

The plan of the article is as follows.
In Sec. \ref{phsd} we review the PHSD approach reminding also the implementation of the retarded electromagnetic fields.
In Sec. \ref{emfields}, we discuss how the transverse components of those fields are distributed in space and time in proton-gold collisions at top RHIC energy.
Our results on particle distributions and flow harmonics are presented respectively in Sec. \ref{pAu} and \ref{collectivity}, highlighting in particular the role of the electromagnetic fields.
Finally, in Sec. \ref{conclusions} we draw our conclusions.

\section{Particle and electromagnetic field evolution in the PHSD approach}
\label{phsd}

The dynamical evolution of heavy-ion collisions as well as small colliding systems at relativistic energy is described by means of the Parton-Hadron-String Dynamics (PHSD) approach, which is a covariant dynamical model for strongly interacting many-body systems formulated on the basis of generalized transport equations, which are derived from the off-shell Kadanoff-Baym equations for non-equilibrium Green functions in phase-space representation \cite{Juchem:2003bi, Juchem:2004cs, Cassing:2009vt, Cassing:2007yg, Cassing:2008sv, Cassing:2008nn}.
The Kadanoff-Baym theory treats the field quanta in terms of dressed propagators with complex self-energies, whose real and imaginary parts can be related respectively to mean-field potentials and particle widths \cite{Cassing:2008nn}.
The off-shell transport equations fully governs the time evolution of the system both in the partonic and in the hadronic phase, once the proper complex self-energies of the degrees of freedom are known \cite{Juchem:2003bi, Juchem:2004cs, Cassing:2008nn}.
See Ref.~\cite{Cassing:2008nn} for a review on off-shell transport theory. 

In the beginning of the nuclear collision, primary nucleon-nucleon hard inelastic scatterings between the two impinging nuclei lead to the formation of color-neutral strings described by the FRITIOF model \cite{Pi:1992ug, Andersson:1992iq} based on the Lund string fragmentation picture.
These strings fragment into ``pre-hadrons'', which are baryons and mesons within their formation time $\tau_f$ (taken to be 0.8 fm/$c$ in their rest frame) and do not interact with the surrounding medium, and into ``leading hadrons'', which are the fastest residues of the string ends and can reinteract with other hadrons with reduced cross sections in line with quark-counting rules.
The fate of the pre-hadrons is determined by the local energy density. If it is above the critical energy density of the deconfinement transition, which is taken to be $\epsilon_c=0.5$ GeV/fm$^3$, pre-hadrons dissolve in massive quarks, antiquarks and gluons plus a mean-field potential. The properties of these colored quasi-particles are determined by the Dynamical Quasi-Particle Model (DQPM) \cite{Cassing:2009vt}, which defines the parton spectral functions, i.e. masses $M_{q,g}(\epsilon)$ and widths $\Gamma_{q,g}(\epsilon)$, and self-generated repulsive mean-field potentials $U_{q,g}(\epsilon)$. Within the DQPM the local energy density $\epsilon$ is related through the lQCD equation of state to the temperature $T$ in the local cell.
In the DQPM model, the temperature-dependent effective masses and widths of quarks, antiquarks and gluons are fitted to the lQCD thermodynamic quantities, such as energy density, pressure and entropy density. Moreover, the partonic interaction rates derived from the DQPM give rise to transport properties of the hot QGP that are in line with the lQCD results; indeed, the shear and bulk viscosities as well as electric conductivity agree to a good extent to the corresponding transport coefficients computed on the lattice \cite{Ozvenchuk:2012kh, Cassing:2013iz}. 
The transition from the partonic to hadronic degrees of freedom is described by dynamical hadronization, which is modeled by means of covariant transition rates for the fusion of quark-antiquark pairs to mesonic resonances and three quarks or antiquarks to baryonic states \cite{Cassing:2009vt, Bratkovskaya:2011wp}.
Thanks to the off-shell nature of both partons and hadrons, the hadronization process fulfil flavor-current conservation, color neutrality as well as energy-momentum conservation and obey the second law of thermodynamics of total entropy increase.
In the hadronic phase, i.e., for energies densities below the critical energy density $\epsilon_c$, the PHSD approach is equivalent to the Hadron-Strings Dynamics (HSD) model \cite{Cassing:1999es}.
PHSD has demonstrated its capability to provide a good description of nucleus-nucleus collisions from the lower superproton-synchrotron (SPS) to the top LHC energies for bulk observables and collective flows \cite{Cassing:2009vt, Bratkovskaya:2011wp, Konchakovski:2011qa, Konchakovski:2012yg, Konchakovski:2014fya} as well as for electromagnetic probes \cite{Linnyk:2011hz, Linnyk:2012pu, Linnyk:2015tha, Linnyk:2015rco}.

The PHSD approach has been extended by including for all binary partonic channels differential off-shell scattering cross sections as a function of temperature $T$ and baryon chemical potential $\mu_B$, on the basis of the effective propagators and couplings from the DQPM that is matched to reproduce the QGP equation of state computed on the lattice (PHSD5.0) \cite{Moreau:2019vhw}.
Nevertheless, in this work we have used the default version (PHSD4.0).

PHSD includes the dynamical formation and evolution of the retarded electromagnetic field (EMF) and its influence on quasi-particle dynamics \cite{Voronyuk:2011jd}. In order to obtain a consistent solution of particle and field evolution equations, the off-shell transport equation are supplemented by the Maxwell equations for the electric field ${\bm E}$ and the magnetic field ${\bm B}$.
Expressing the fields in terms of the scalar potential $\Phi$ and the vector potential ${\bm A}$:
\begin{equation}\label{fields_potentials}
{\bm E}=-\nabla\Phi-\frac{\partial{\bm A}}{\partial t}, \qquad {\bm B}=\nabla\times{\bm A},
\end{equation}
one obtains a wave equation for the potentials whose solution for an arbitrarily moving point-like charge is given by the Liénard-Wiechert potentials. Inserting them into Eq.~\eqref{fields_potentials}, the electric and magnetic fields generated by a point-like source with charge $e$ at position ${\bm r}(t)$ travelling at velocity ${\bm v}(t)$ are given by
\begin{equation}\label{Elw}
{\bm E}({\bm r},t)=\dfrac{e}{4\pi}
\left\lbrace
\dfrac{{\bm n}-{\bm\beta}}{\kappa^3\gamma^2R^2}
+\dfrac{{\bm n}\times\left[\left({\bm n}-{\bm\beta}\right)\times\dot{{\bm \beta}}\right]}{\kappa^3cR}
\right\rbrace_{\mathrm{ret}}
\end{equation}
\begin{equation}\label{Blw}
{\bm B}({\bm r},t)=\left\lbrace  {\bm n}\times{\bm E}({\bm r},t)  \right\rbrace_{\mathrm{ret}}
\end{equation}
where ${\bm R}={\bm r}-{\bm r}'$ with ${\bm r}'\equiv{\bm r}(t')$, ${\bm n}={\bm R}/R$, ${\bm\beta}={\bm v}/c$, $\dot{{\bm\beta}}=\mathrm{d}{\bm\beta}/\mathrm{d}t$ and $\kappa=1-{\bm n}\cdot{\bm\beta}$; the subscript ``ret'' means that the quantities inside the braces have to be evaluated at the times $t'$ that are solutions of the retardation equation $t'-t+{\bm R}(t')/c=0$.
We see from the previous equation that retarded electromagnetic fields from moving charges divide themselves naturally into two contributions: the first term represents ``velocity fields'', which are independent of acceleration and are essentially elastic Coulomb fields varying for large $R$ as $R^{-2}$; the second term describes ``acceleration fields'', which depend linearly on the acceleration and are intepreted as radiation fields falling off for large distances as $R^{-1}$ \cite{Landau:1975}. 
Neglecting the acceleration $\dot{{\bm \beta}}$ in Eqs.~\eqref{Elw}-\eqref{Blw} and considering that in a nuclear collision the total electric and magnetic field are a superposition of the fields generated from all moving charges, one obtains the final formulas implemented in the PHSD code:
\begin{equation}
e{\bm E}({\bm r},t)=\sum_i
\dfrac{\mathrm{sgn}(q_i)\alpha_{em}{\bm R}_i(t)(1-\beta_i^2)}{\left\lbrace\left[{\bm R}_i(t)\cdot{\bm\beta}_i\right]^2+R_i(t)^2\left(1-\beta_i^2\right)\right\rbrace^{3/2}},
\end{equation}
\begin{equation}
e{\bm B}({\bm r},t)=\sum_i
\dfrac{\mathrm{sgn}(q_i)\alpha_{em}{\bm\beta_i}\times{\bm R}_i(t)(1-\beta_i^2)}{\left\lbrace\left[{\bm R}_i(t)\cdot{\bm\beta}_i\right]^2+R_i(t)^2\left(1-\beta_i^2\right)\right\rbrace^{3/2}},
\end{equation}
where the sum over $i$ runs over all particles with charge $q_i$ and $\alpha_{em}=e^2/4\pi\simeq1/137$ is the electromagnetic fine-structure constant.
The quasiparticle propagation in the electromagnetic field is determined by the Lorentz force:
\begin{equation}\label{lorentz}
\left(\dfrac{\mathrm{d}{\bm p}_i}{\mathrm{d}t}\right)_{em}=q_i\left({\bm E}+{\bm\beta}_i\times{\bm B}\right)
\end{equation}

It is not clear a priori which is the response to the electromagnetic field of a particle under formation (as previously mentioned a formation time $\tau_f$ for newly produced particles is considered in PHSD). The suppression of the electromagnetic coupling to the charge of unformed hadrons and partons can be called as the “inverse LPM” (iLPM) effect \cite{Toneev:2016bri}. The iLPM effect is incorporated in PHSD as a time delay $\tau_f^{em}$ in the interaction of the electromagnetic field with the charged degrees of freedom. Since the charge is a conserved quantity, preformed charged particles should sense the electromagnetic field long before being completely formed; hence in PHSD $\tau_f^{em}=\tau_f/10$ is assumed. See Ref.~\cite{Toneev:2016bri} for a detailed explanation and discussion of the iLPM effect.

Solving the generalized transport equations with the inclusion of the Lorentz force term accounts to have a consistent description of the dynamical evolution of the strongly-interacting many-body system with the self-generated electromagnetic fields.

\section{Electromagnetic field distributions in p+Au collisions}
\label{emfields}

\begin{figure*}[h!]
\centering
{\includegraphics*[trim={30 5 0 15},width=0.8\columnwidth]{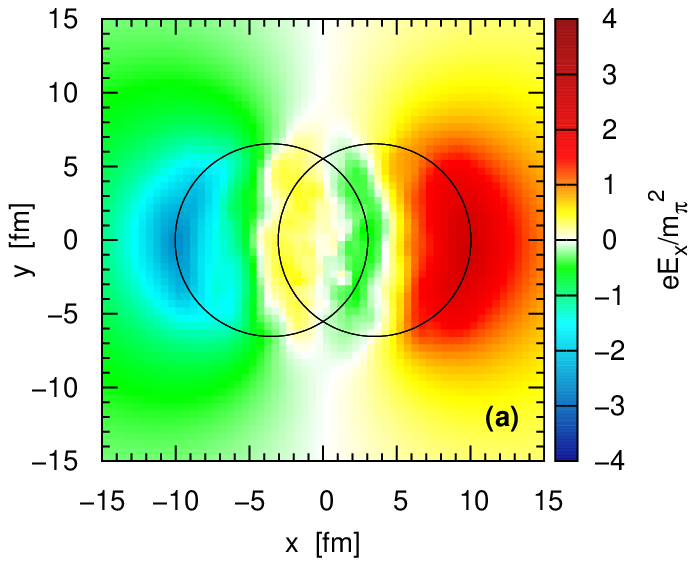}
\includegraphics*[trim={30 5 0 15},width=0.8\columnwidth]{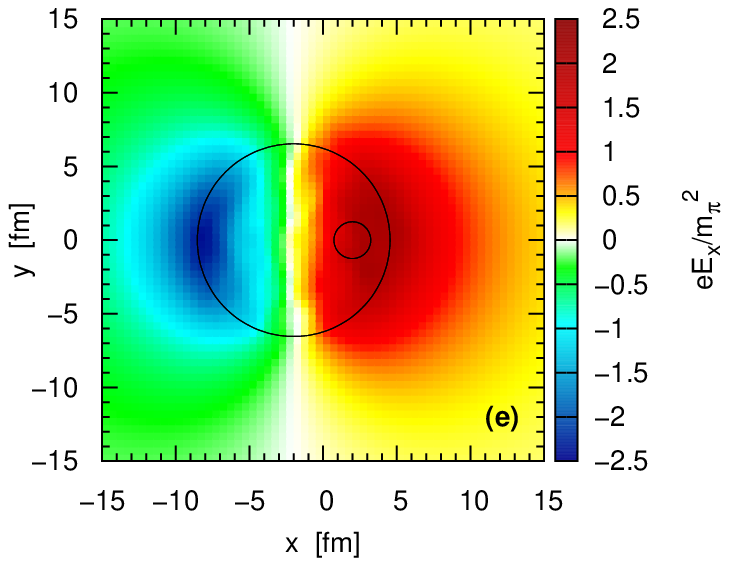}}
{\includegraphics*[trim={30 5 0 15},width=0.8\columnwidth]{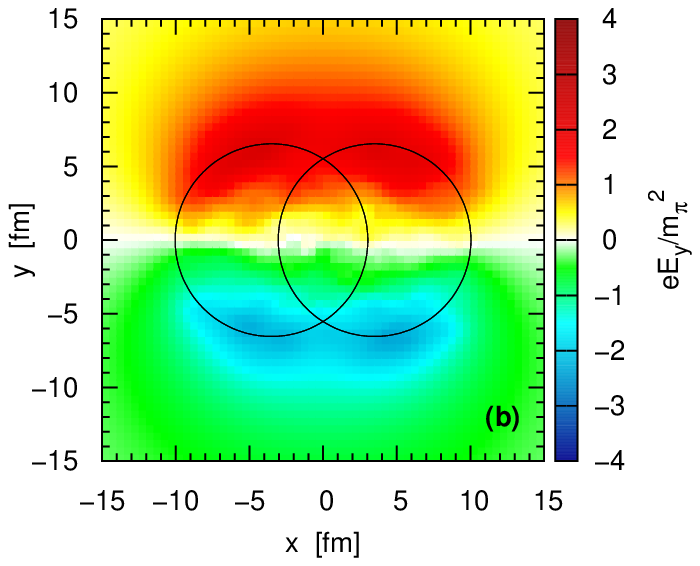}
\includegraphics*[trim={30 5 0 15},width=0.8\columnwidth]{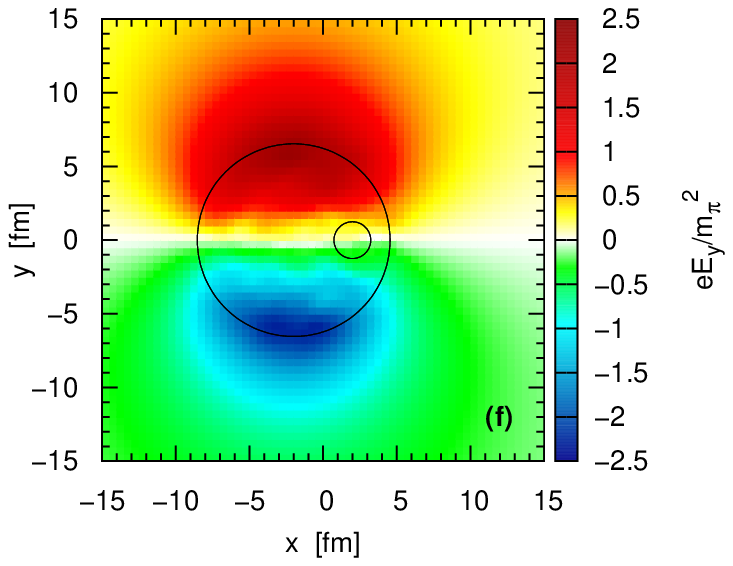}}
{\includegraphics*[trim={30 5 0 15},width=0.8\columnwidth]{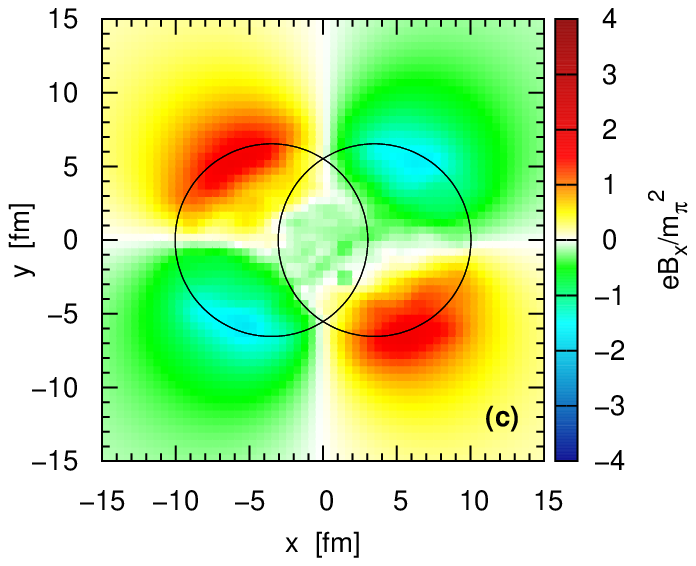}
\includegraphics*[trim={30 5 0 15},width=0.8\columnwidth]{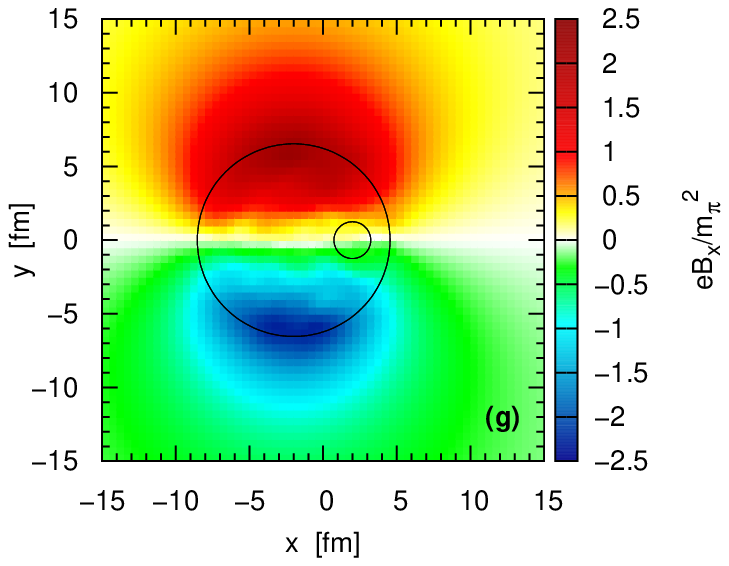}}
{\includegraphics*[trim={30 5 0 15},width=0.8\columnwidth]{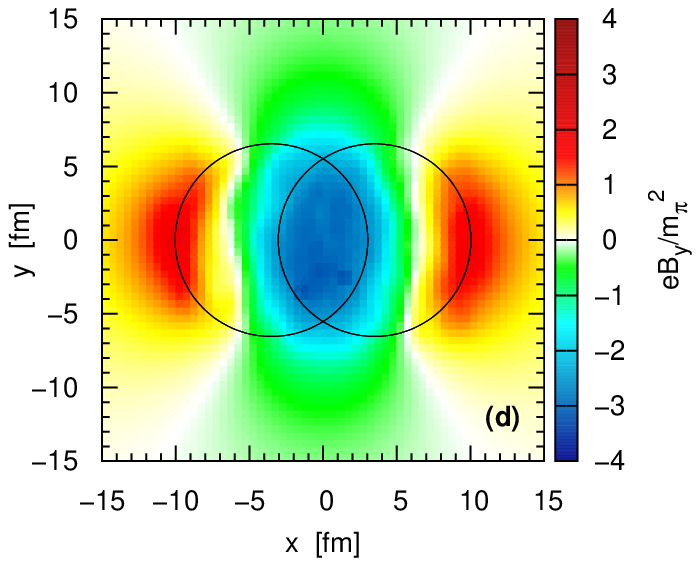}
\includegraphics*[trim={30 5 0 15},width=0.8\columnwidth]{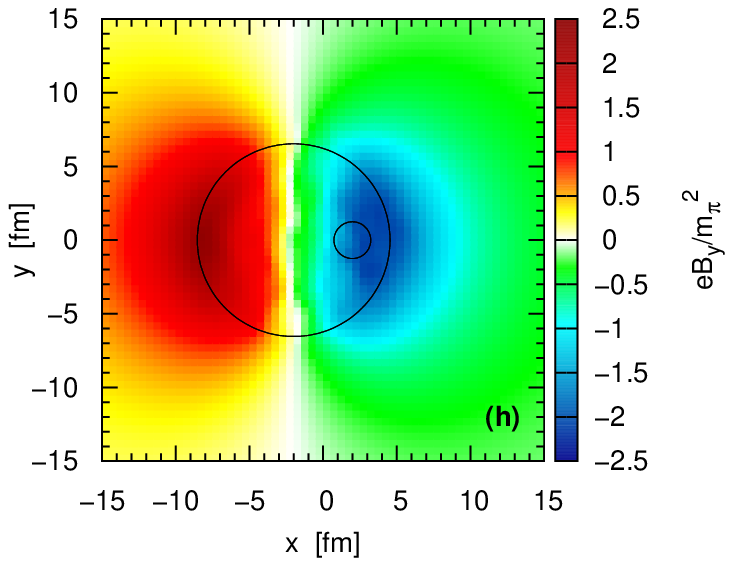}}
\\[\baselineskip]
(a)--(d): \emph{Au+Au, $b=7$ fm}~~~~~~~~~~~~~~~~~~~~~~~~~~~~~~~~(e)--(h): \emph{p+Au, $b=4$ fm}~~~~~~~
\caption{(Color online) Distribution of the electromagnetic fields $E_x$, $E_y$, $B_x$, $B_y$ in the transverse plane at $z=0$ at the maximum overlap time of Au+Au collisions with impact parameter $b=7$ fm (panels (a)--(d)) and p+Au collisions with $b=4$ fm (panels (e)--(h)) at $\sqrt{s_{NN}}=200$ GeV. Circles roughly corresponding to the position of the proton and gold nuclei are drawn to guide the eye.}
\label{fields_AuAu_pAu}
\end{figure*}

In this section we discuss and compare the distribution in strength and direction of the electromagnetic fields for off-central Au+Au and p+Au collisions.
The field distributions changes remarkable going from symmetric collisions (e.g. Au+Au \cite{Voronyuk:2011jd}), where the two nuclei have the same size and number of proton, through asymmetric collisions (e.g. Cu+Au \cite{Voronyuk:2014rna}), where the two nuclei have different size and atomic number, to proton-induced reactions (e.g. p+Au), when the field distributions is basically the one produced by the heavy nucleus.

In Fig.~\ref{fields_AuAu_pAu} we show the transverse components of the electromagnetic field produced at top RHIC energy in Au+Au collisions at $b=7$ fm (panels (a)--(d)) and p+Au collisions at $b=4$ fm (panels (e)--(h)).
The field strength are computed at the time when both nuclear centers are in the same transverse plane, namely at the maximum overlap time of the collision.
In all panels circles roughly corresponding to the size of proton and gold nuclei are drawn in order to guide the eye highlighting the interaction area.
It has been pointed out \cite{Skokov:2009qp, Voronyuk:2011jd} that in symmetric nucleus-nucleus collisions the electromagnetic field produced in the early stage of the collision is dominated by the magnetic field along the $y$ direction, i.e. orthogonal to the reaction plane. We see from Fig.~\ref{fields_AuAu_pAu} (d) that this component reaches value $\left|eB_y\right|\simeq 4\,m_{\pi}^2$ whereas the $x$-component of the magnetic field as well as the electric fields components are nearly vanishing or very low.
In Ref.~\cite{Voronyuk:2014rna}, within the PHSD framework, it has been shown that in Cu+Au collisions a significant electric field $E_x$ directed from the heavier gold nucleus towards the lighter copper nucleus is generated in the central region of the overlap area of the collision. This is due to the different number of protons in the two colliding nuclei.
In p+Au collisions the initial electromagnetic field distributions correspond basically to the one produced by the gold nucleus moving at velocity close to the speed of light and do not depend significantly on the impact parameter of the collision. Nevertheless, the point in which the proton hits the gold nuclei has a strong impact on the field magnitude filled by it.    
From Fig.~\ref{fields_AuAu_pAu} (e) and (f) we see that for non-central p+Au collisions the electric field produced is strongly asymmetric inside the overlap area; in this region, for collision at $b=4$ fm, both $\left|eB_y\right|$ (h) and $\left|eE_x\right|$ (e) reaches values $\simeq 2\,m_{\pi}^2$, while the other electromagnetic field components are close to zero.

\begin{figure}[t!]
\centering
\includegraphics[width=\columnwidth]{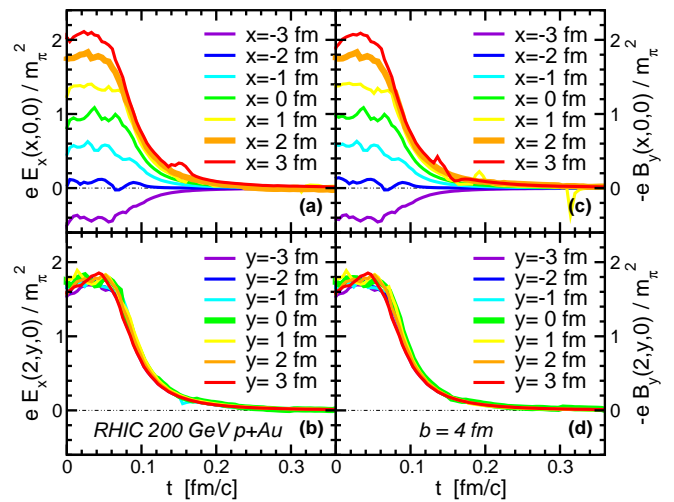}
\caption{(Color online) Time evolution of the event-averaged electromagnetic field components $E_x$ (panels (a) and (b)) and $B_y$ (panels (c) and (d)) for p+Au collisions at $\sqrt{s_{NN}}=200$ GeV with impact parameter $b=4$. The different lines correspond to different values of $x\in[-3,3]$ fm at $y=z=0$ fm (panels (a) and (c)) and different values of $y\in[-3,3]$ fm at $x=b/2=2$ fm and $z=0$ fm (panels (b) and (d)).}
\label{emf_t}
\end{figure}

From Fig.~\ref{emf_t} one can better see how the maximum values of $\left|E_x\right|$ (panels (a) and (b)) and $\left|B_y\right|$ (panels (c) and (d)) changes for p+Au collisions at $b=4$ fm moving in the transverse plane respect to the centre of the overlap area $\lbrace x,y\rbrace=\lbrace 2\;\mathrm{fm},0\;\mathrm{fm}\rbrace$.
Moreover, this plot gives information also on the temporal evolution of the two electromagnetic field components, showing that both $\left|B_y\right|$ and $\left|E_x\right|$ decrease very fast, becoming close to zero after $\sim0.25$ fm/$c$ from the first nucleon-nucleon collisions corresponding to $t=0$ fm/$c$.
While moving in the $x-$direction the value of $\left|E_x\right|(x,0,0)$ and $\left|B_y\right|(x,0,0)$ changes of about 20-25\% in 1 fm, spanning the $y-$direction $\left|E_x\right|(2,y,0)$ and $\left|B_y\right|(2,y,0)$ remain almost homogeneous in a region of at least 6 fm.
Hence, for non-central collisions there is a wide area around the point where the proton hits the nucleus in which not only the magnetic field but also the electric field is intense, even though they last only for a small fraction on fm/$c$. The electromagnetic field acts as an accelerator on charges that are present during this time according to the Lorentz force Eq.~\eqref{lorentz}.
However, considering that in a relativistic nuclear collisions the evolution of the fireball is dominated by the longitudinal expansion (at least in the early stage) and taking into account strengths and directions of the electromagnetic fields, it turns out that the electric and the magnetic part of the Lorentz force push the charges in opposite directions, thus ending up with a partial cancellation of the corresponding momentum increments of charged particles in each cell \cite{Toneev:2011aa}.
Realistic simulations are needed in order to extract the dynamical influence of the electromagnetic field on final observables.
The cancellation of the electric and magnetic pushes is very strong in symmetric heavy ion collisions \citep{Voronyuk:2011jd}.
We will see in Sect.~\ref{v1_results} that in the strongly asymmetric p+Au systems, due to the high value of $E_x$, the electromagnetic field has a remarkable effect on the directed flow of mesons.

\section{Properties of p+Au collisions}
\label{pAu}

\subsection{Particle distributions and centrality selection}
\label{centr_select}

\begin{figure}[h!]
\centering
\includegraphics[trim={0 10 0 35},width=0.95\columnwidth]{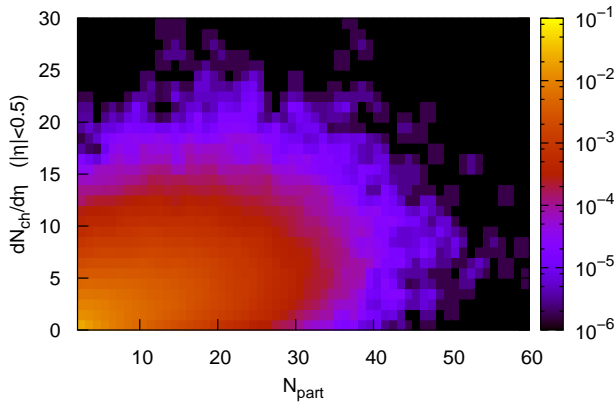}
\caption{(Color online) Event probability as a function of the number of participants and the number of charged particles at midrapidity ($\left|\eta\right|<0.5$) for minimum bias p+Au collisions at $\sqrt{s_{NN}}=200$ GeV.}
\label{events_Npart_Nch}
\end{figure}

\begin{figure*}[t!]
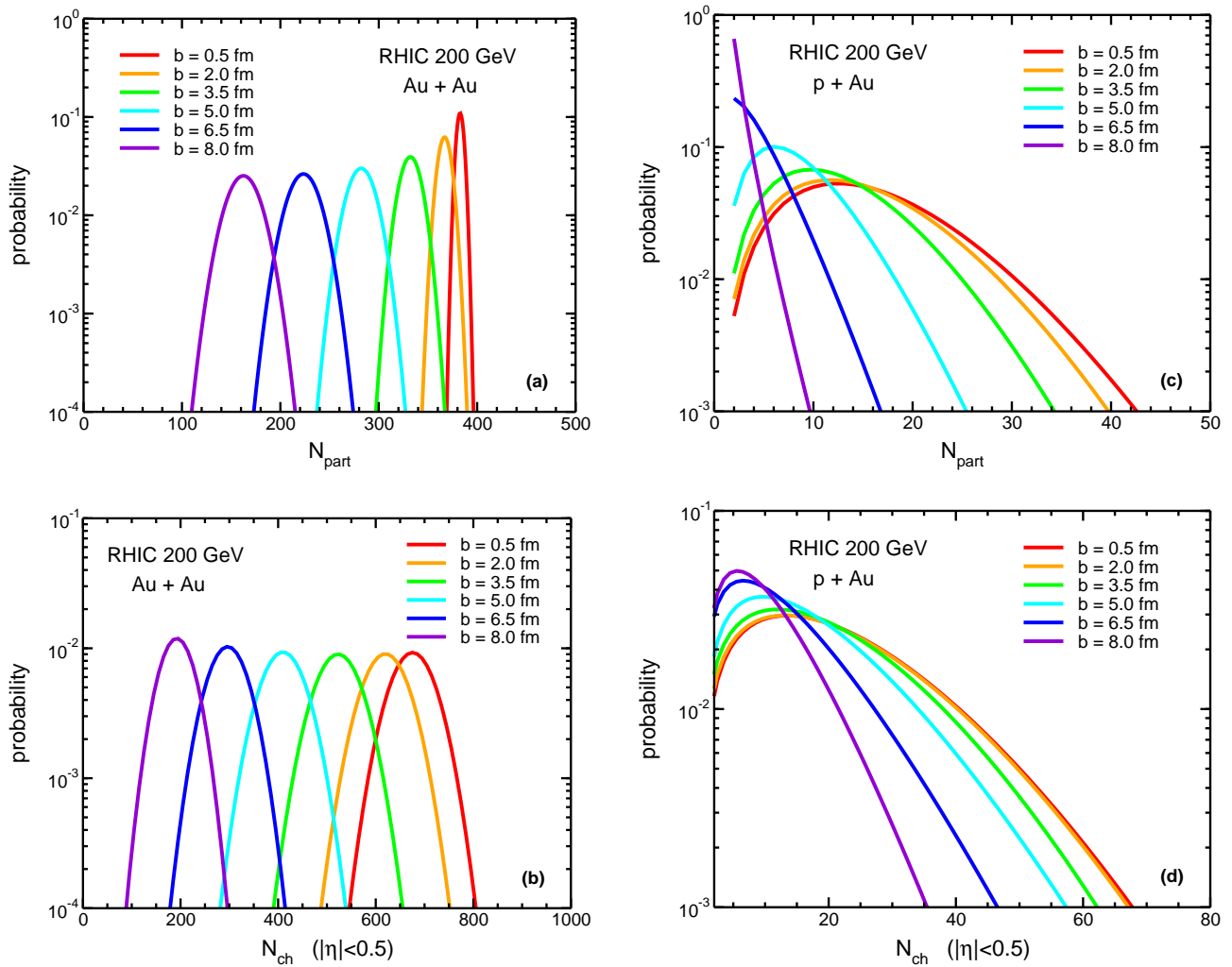

\centering
{\includegraphics[width=0.95\columnwidth]{fig_prob_Npart_b_AuAu.eps}\qquad
\includegraphics[width=0.95\columnwidth]{fig_prob_Npart_b_pAu.eps}}
\\[\baselineskip]
{\includegraphics[width=0.95\columnwidth]{fig_prob_Nch_b_AuAu.eps}\qquad
\includegraphics[width=0.95\columnwidth]{fig_prob_Nch_b_pAu.eps}}
\caption{(Color online) Event probability as a function of the number of participants (panels (a) and (c)) and of the number of charged particles at midrapidity ((b), (d)) for Au+Au collisions (panels (a) and (b)) and p+Au collisions (panels (c) and (d)) at $\sqrt{s_{NN}}=200$ GeV; in all plots the different curves correspond to different impact parameters from $b=0.5$ fm to $b=8$ fm (lines from right to left).}
\label{events_b}
\end{figure*}

Experimentally, the collision geometry cannot be controlled and initial state quantities, such as the impact parameter $b$, the number of nucleon-nucleon collisions $N_{coll}$ and the number of participating nucleons $N_{part}$, cannot be accessed in a direct way. The (MC-)Glauber model has been widely used to describe the collision geometry, estimating the initial spatial distribution of nucleons in the transverse plane, and to connect experimental observables with the theoretically evaluated $b$, $N_{coll}$, $N_{part}$ \cite{Miller:2007ri}.
In heavy-ion collisions, different geometries -- different $b$ -- correspond to different $N_{coll}$ and $N_{part}$ and the concept of collision geometry is strictly connected to the concept of collision centrality, which characterizes the size of the overlap area between the two nuclei. Nevertheless, there is not a unique definition of centrality and many observables can be used to its determination, e.g., the number of charged particle $N_{ch}$ produced at midrapidity. Moreover, due to fluctuations in particle production, there are fluctuations between initial and final state quantities: for example events with the same $N_{part}$ may correspond to different amount of $N_{ch}$ and viceversa. These centrality fluctuations lead in turn to an uncertainty in the interpretation of experimental measurements and its centrality dependence \cite{Jeon:2003gk, Skokov:2012ds, Luo:2013bmi, Zhou:2018fxx}.

In the case of small collision systems, such us proton-induced collisions, the centrality fluctuations become so huge that the concept itself of centrality changes. Indeed, it is not possible to correlate clearly the collision geometry, the size of the fireball and the amount of produced particle. The centrality determination in small systems is based on the measured particle multiplicity and loses its strong link to the collision impact parameter \cite{Adare:2013nff, Oppedisano:2014vja, Aidala:2016vgl, Aidala:2017pup}.
This implies new efforts from theoretical simulations in order to compare their results to experimental measurements and to make predictions that can be easily verified.

Within the PHSD approach we can choose between two different procedures to simulate proton-nucleus collisions: by fixing the value of the impact parameter we can study and comprehend in a more clear way the influence of the collision geometry and phenomena strictly related to it (e.g., the generation of electromagnetic fields), whereas reproducing minimum bias collisions with impact parameter values randomly distributed according with the correct geometric probability allow us a more direct understanding of experimental measurements (e.g., the particle distributions for given centrality class).

In A-A collisions there is a clear scaling behaviour of particle production with the participant number, $dN_{ch}/d\eta\propto N_{part}^\alpha$, with the parameter $\alpha$ depending on the collision energy \cite{Adler:2004zn, Abelev:2013qoq}. 
In small system there is still a correlation between the charged particle multiplicity at midrapidity and the participant number, but with large dispersion in both quantities respect to A-A collisions  \cite{Konchakovski:2014wqa}. This is shown for p+Au collisions at $\sqrt{s_{NN}}=200$ GeV in Fig.~\ref{events_Npart_Nch}, where the probability distribution in the number of participant $N_{part}$ and the number of charged particles $N_{ch}$ produced at $\left|\eta\right|<0.5$ obtained with minimum bias PHSD simulations is plotted.
The reason of this huge dispersion is that the proton can hit a very different number of nucleons inside the gold nucleus, up to a maximum of about 50 nucleons, and for a given number of participating nucleons the number of charged particle produced can vary widely, from zero to about 20 at midrapidity.

In order to better understand how these quantities correlates with the geometry of the collision and how this connection changes from large to small colliding systems, we show in Fig.~\ref{events_b} the event probability of the number of participants (panels (a) and (c)) and the number of charged particles in $\left|\eta\right|<0.5$ (panels (b) and (d)) computed with PHSD for six different values of impact parameters, from $b=0.5$ fm to $b=8.0$ fm, for Au+Au collisions at $\sqrt{s_{NN}}=200$ GeV (panels (a) and (b)) and for p+Au collisions at the same energy (panels (c) and (d)).
While for Au+Au collisions the probability distribution of $N_{part}$ and $N_{ch}$ at midrapidity is close to a gaussian around a mean value which increases for decreasing $b$, providing a good correlation between the three quantities, for p+Au systems multiplicity fluctuations mix events from very different impact parameters and collisions with $b\lesssim2$ fm give almost the same results for what concerns the participant number and the produced multiplicity at midrapidity; the fluctuations of these quantities at fixed impact parameter are better described by a gamma distribution than by a gaussian one in the case of proton-nucleus collisions \cite{Rogly:2018ddx}.

For the results presented in the next sections, we have selected centrality in our minimum bias simulations computing the number of charged particle in the pseudorapidity window $\left|\eta\right|<2$ and dividing the events in multiplicity bins.

\subsection{Rapidity distributions and transverse momentum spectra}

\begin{figure}[h!]
\centering
\includegraphics[width=\columnwidth]{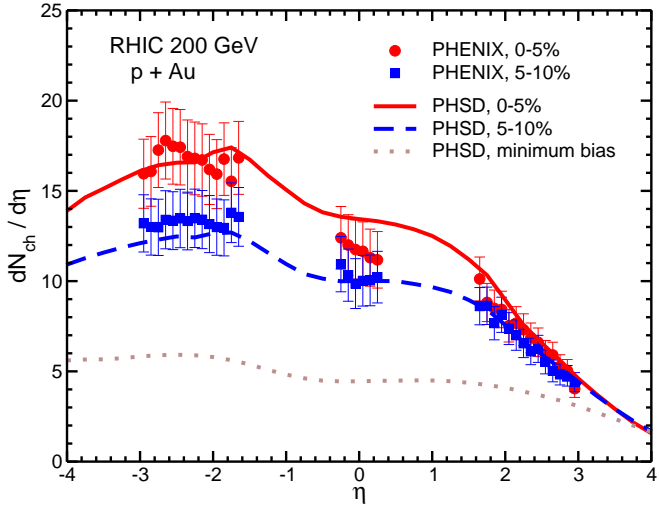}
\caption{(Color online) Pseudorapidity distribution of charged particles for p+Au collisions at $\sqrt{s_{NN}}=200$ GeV computed with PHSD simulations for minimum bias events (dotted brown curve) and for the 5\% (solid red curve) and 5-10\% (dashed blue curve) most central collisions; for the latter two results the corresponding experimental data from the PHENIX Collaboration \cite{Adare:2018toe} are shown for comparison (red circles and blue squares respectively).}
\label{rap_distr_ch}
\end{figure}

\begin{figure}[h!]
\centering
\includegraphics[width=\columnwidth]{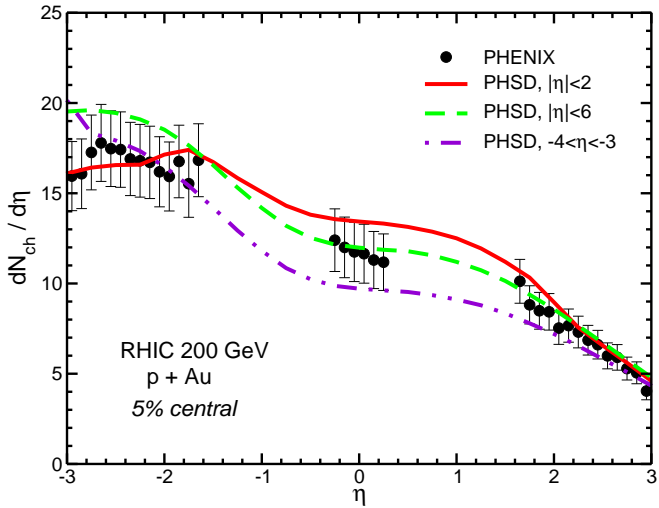}
\caption{(Color online) Impact on charged particle pseudorapidity distribution of using different $\eta$-range for select centrality bins from minimum bias p+Au collisions at $\sqrt{s_{NN}}=200$ GeV; the red, green and violet curves correspond the results for the 5\% most central events obtained selecting centrality with the pseudorapidity ranges $\left|\eta\right|<2$, $\left|\eta\right|<6$ and $-4<\eta<-3$ respectively (see text for more details). The black dots are the experimental data from the PHENIX Collaboration \cite{Adare:2018toe} for the 5\% centrality class.}
\label{rap_distr_ch_cscompare}
\end{figure}

In Fig.~\ref{rap_distr_ch} we plot the charged-particle pseudorapidity density for the 0-5\% (solid red curve) and 5-10\% (dashed blue curve) most central collisions along with the minimum bias result (dotted brown line) for the p+Au system at $\sqrt{s_{NN}}=200$ GeV.
In agreement with the corresponding data from the PHENIX Collaboration \cite{Adare:2018toe} (red circles and blue squares), the charged-particle distributions are asymmetric in pseudorapidity $\eta$ and present an enhancement at backward rapidity, i.e., in the Au-going side.
Moreover, they vary strongly with centrality, with an increasing asymmetry between the proton-going and Au-going directions as the collisions become more central; indeed, the minimum bias result is much flatter respect to the most central collisions, and the 0-5\% bin is more asymmetric respect to the 5-10\% one.

We have investigated how the use of different pseudorapidity regions to compute charged particles and then divide the events in centrality classes affects the final results for pseudorapidity distributions of charged particles. This is shown in Fig.~\ref{rap_distr_ch_cscompare} for the 5\% most central collision, comparing the result in Fig.~\ref{rap_distr_ch} (solid red curve) with that obtained computing charged particles in $\left|\eta\right|<6$ (dashed green curve) and $-4<\eta<-3$ (dot dashed violet line).
The analysis suggests that the method adopted to define centrality and even the specific choices of the method -- e.g. the pseudorapidity range used to compute charged particles -- may have a significant impact on the centrality dependence of the measured $dN_{ch}/d\eta$ distribution.
In particular, the latter shows a more pronounced asymmetry if we use the range $-4<\eta<-3$ or $\left|\eta\right|<6$ respect to $\left|\eta\right|<2$, meaning that some contribution to the asymmetry in rapidity densities comes from the use for centrality determination of $\eta$-regions far from midrapidity, where there is the larger difference in particle production between forward and backward rapidity. 
Consequently, we expect a similar impact of these choices also on rapidity distributions of identified particles.

\begin{figure}[t!]
\centering
\includegraphics[width=\columnwidth]{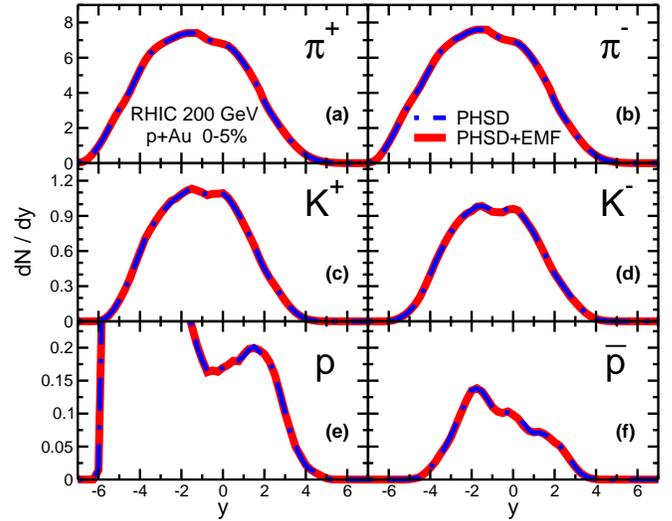}
\caption{(Color online) Rapidity distributions of identified particles for 5\% central p+Au collisions at $\sqrt{s_{NN}}=200$ GeV obtained with PHSD simulations with (solid red lines) and without (dot-dashed blue lines) the inclusion of electromagnetic fields.}
\label{rap_distr_id}
\end{figure}

\begin{figure}[h!]
\centering
\includegraphics[width=\columnwidth]{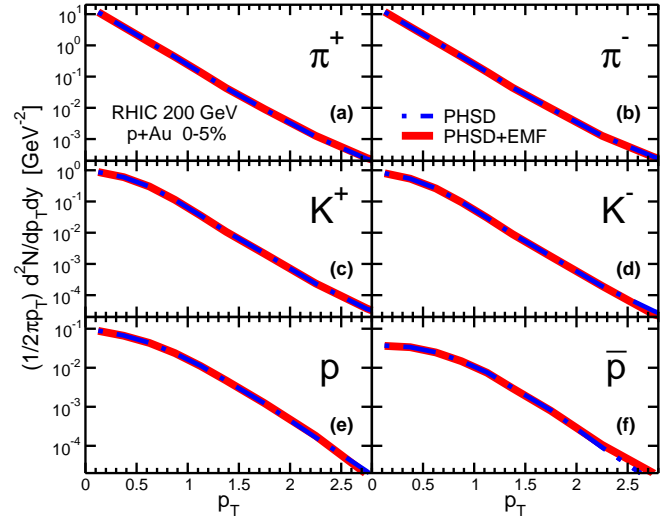}
\caption{(Color online) Transverse momentum spectra of identified hadrons for 5\% central p+Au collisions at $\sqrt{s_{NN}}=200$ GeV obtained with PHSD simulations with (solid red lines) and without (dot-dashed blue lines) the inclusion of electromagnetic fields.}
\label{spectra_id}
\end{figure}

The rapidity distributions of $\pi^+$, $\pi^-$, $K^+$, $K^-$ for 5\% central collisions are displayed in Fig.~\ref{rap_distr_id}.
We notice again that rapidity distributions are asymmetric in rapidity $y$ and present an enhancement in the Au-going side ($y<0$).
We have checked whether the electromagnetic fields produced in p+Au collisions could affect the final rapidity densities comparing simulations with and without the inclusion of those fields, represented respectively by solid red lines and dot-dashed blue lines and the in Fig.~\ref{rap_distr_id}. We find no change in the hadron distributions of 5\% central collisions and we have checked that any influence is seen in other centrality classes and in minimum bias collisions as well.
The same applies to the transverse momentum spectra of hadrons at midrapidity $|\eta|<0.5$ shown in Fig.~\ref{spectra_id} for the same most central bin. There is practically no difference in the PHSD results with and without the electromagnetic field for pions, kaons, protons and antiprotons.
Nevertheless, our results on rapidity densities and momentum spectra in this strongly asymmetric system can be considered as predictions for the production of the most abundant hadron species at top RHIC energy.

\begin{figure*}[hbt!]
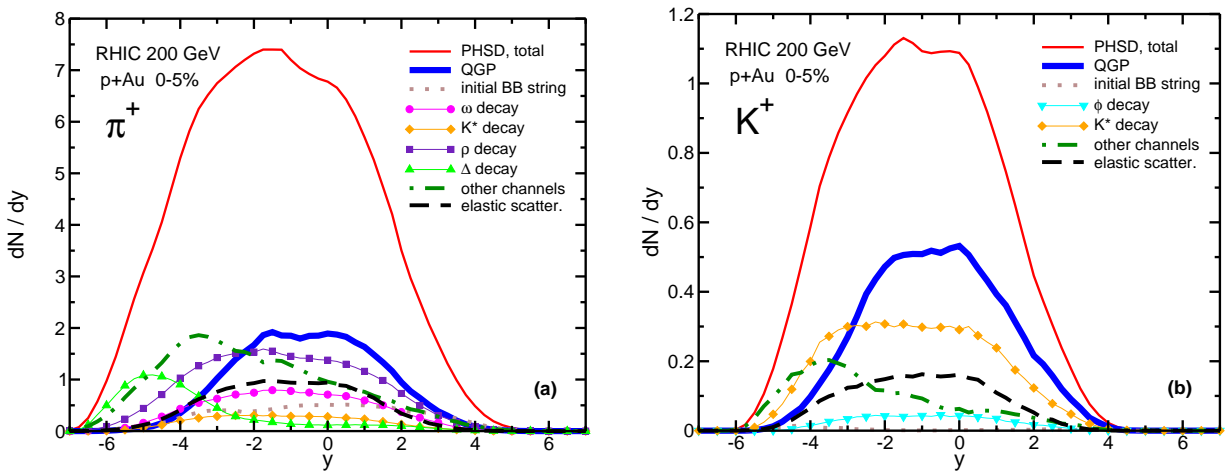

\centering
{\includegraphics[width=0.9\columnwidth]{fig_dN_y_0-5_pi_history.eps}\qquad
\includegraphics[width=0.9\columnwidth]{fig_dN_y_0-5_K_history.eps}}
\caption{(Color online) Channel decomposition of the rapidity distributions of $\pi^+$ (a) and $K^+$ (b) for 5\% central p+Au collisions at $\sqrt{s_{NN}}=200$ GeV calculated with PHSD.}
\label{dNdy_channels}
\end{figure*}

In Fig.~\ref{dNdy_channels} we show the channel decomposition of the rapidity distributions of $\pi^+$ (a) and $K^+$ (b); the results for $\pi^-$ and $K^-$ are similar to those of the corresponding antiparticle, hence an explicit representation is discarded.
Regarding pion production, we see that remarkable contributions come from QGP hadronization (solid thick blue line), decay of resonances ($\Delta$, $\rho$, $\omega$, $K^*$), initial baryon-baryon strings (dotted brown line) and other channels (dot-dashed green line) such as meson-meson and meson-baryon strings. Among the resonances, the $\rho$ vector meson (violet squares) constitutes the main production channel of pions in the central rapidity region, while pions in the target region comes primarily from decay of the $\Delta$ baryon (light-green triangles).
The dashed black line represent pions whose last interaction is an elastic scattering with other hadrons, and thus the information about their production channel is lost.
For what concerns kaons, we point out that at midrapidity, besides a big contribution from the decay of $K^*$ resonance (orange diamonds), there is a large production directly from QGP hadronization (solid thick blue line); this is an interesting difference with respect to A--A collisions in which the kaons created by $K^*$ decay are about twice those generated directly from QGP \cite{Ilner:2016xqr}.
The other curves represent kaons coming from $\phi$ decay (light-blue triangles), initial baryon-baryon strings (dotted brown line) and other production channels (dot-dashed green line) such as hadronic strings; the dashed black line indicates kaons which lastly participate in elastic scatterings.
It is interesting to note that for both pions and kaons there is a noticeable amount of particle escaping from the medium just after production, without undergoing further rescattering.

It may be interesting to look at the highest multiplicity events that we have analized in this section because the charged distribution asymmetry coupled with the asymmetry of the distribution of $E_x$ could play a bigger role in other observables. Hence in the following we will focus on the 5\% collision with high multiplicity in order to investigate the emergence of collective patterns and grasp possible effects of the generated electromagnetic fields.

\section{Collectivity in p+Au collisions}
\label{collectivity}

The most direct experimental evidence of the generation of collective flow comes from the observation of anisotropic radial flow in the $x-y$ plane perpendicular to the beam $z-$axis. It is often characterized by the Fourier expansion in momentum space of the azimuthal particle distribution, whose first two coefficients are the directed flow $v_1$ and the elliptic flow $v_2$ respectively given by
\begin{equation}\label{v1}
v_1=\left\langle\cos \phi\right\rangle \equiv \left\langle p_x/p_T \right\rangle,  
\end{equation}
\begin{equation}\label{v2}
v_2=\left\langle\cos 2\phi\right\rangle \equiv \left\langle (p_x^2-p_y^2)/p_T^2 \right\rangle,
\end{equation}
where $\phi$ is the azimuthal angle of the particle (in momentum space) and the brakets indicate average over all events.

In order to take into account the event-by-event flow fluctuations, we compute in each event the $n$-th order event-plane angles $\Psi_n$ from the final-state momentum distribution and the azimuthal anisotropy harmonics with respect to the correspondent event-plane angle; then, the final result is obtained averaging over all events in the centrality class of interest. 
The $n$-th order flow harmonics are given by
\begin{equation}\label{vn_Psin}
v_n^{ \; \left\lbrace \Psi_n \right\rbrace }=\dfrac{\left\langle\cos \left[n\left(\phi-\Psi_n\right)\right]\right\rangle}{\mathrm{Res}(\Psi_n)},
\end{equation}
with the n-th order event-plane angle $\Psi_n$ computed as
\begin{equation}
\Psi_n=\frac{1}{n}\mathrm{atan2}(Q_n^y,Q_n^x),
\end{equation}
being $Q_n^x=\sum_i\cos\left[n\phi_i\right]$ and $Q_n^y=\sum_i\sin\left[n\phi_i\right]$ respectively the $x$ and $y$ projection of the flow vector $Q_n$, where the sum runs over all particles in the chosen pseudorapidity range. Since the finite number of particles produces limited resolution in the determination of $\Psi_n$ -- and this is important especially for small colliding system, such as p-A -- the $v_n$ must be corrected up to what they would be relative to the real reaction plane \cite{Poskanzer:1998yz}; this is done by dividing the observed $v_n$ by the event-plane angle resolution $\mathrm{Res}(\Psi_n)$, which can be computed by means of the three-subevent method that correlate independent determinations of $\Psi_n$ in different pseudorapidity regions \footnote{Denoting with A, B and C three different pseudorapidity regions (subevents), the resolution of the nth-order event-plane angle in the subevent A is given by \cite{Xu:2017tkb}
\begin{equation}
\mathrm{Res}(\Psi_n^A)=\sqrt{\dfrac{\left\langle\cos\left[n\left(\Psi_n^A-\Psi_n^B\right)\right]\right\rangle\left\langle\cos\left[n\left(\Psi_n^A-\Psi_n^C\right)\right]\right\rangle}{\left\langle\cos\left[n\left(\Psi_n^B-\Psi_n^C\right)\right]\right\rangle}}
\end{equation}}.
For the determination of $\Psi_n$ and $\mathrm{Res}(\Psi_n)$ we have used the following three pseudorapidity ranges: $-4<\eta<-3$, $-3<\eta<-1$ and $-0.5<\eta<+0.5$. The values are chosen in order to be similar to those of the detectors used by the PHENIX Collaboration for the determination of the 2nd order event plane  \cite{Adare:2018toe,PHENIX:2018lia,Aidala:2016vgl}.

\begin{figure}[t!]
\centering
\includegraphics[width=\columnwidth]{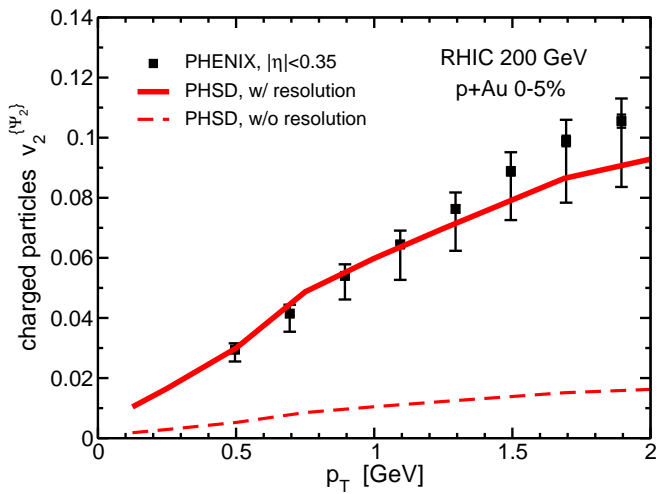}
\caption{(Color online) Comparison of the elliptic flow vs transverse momentum of charged particles at midrapidity for 5\% central p+Au collisions at $\sqrt{s_{NN}}=200$ GeV measured by PHENIX Collaboration \cite{PHENIX:2018lia,Aidala:2016vgl} (black dots) with that obtained with PHSD simulations (red curves): the solid thick line is the $v_2^{\;\left\lbrace\Psi_2\right\rbrace}$ computed with Eq.~\eqref{vn_Psin}, whereas the dashed thin curve is the observed elliptic flow $\left\langle\cos \left[2\left(\phi-\Psi_2\right)\right]\right\rangle$.}
\label{v2_ch_0-5}
\end{figure}

In Fig.~\ref{v2_ch_0-5} we show with a solid red line the PHSD results for the elliptic flow of charged particles respect to the 2nd order event plane $v_2^{\;\left\lbrace\Psi_2\right\rbrace}$ as a function of the transverse momentum for the most central p+Au collisions at $\sqrt{s_{NN}}=200$ GeV. For comparison we plot the experimental data from PHENIX Collaboration \cite{PHENIX:2018lia,Aidala:2016vgl} (black dots).
The resolution of the 2nd order event-plane angle in 5\% central p+Au collisions determined by PHENIX Collaboration is $\mathrm{Res}(\Psi_2^{FVTX-S})=0.171$ \cite{Aidala:2016vgl}, where the FVTX-S detector covers the pseudorapidity range $-3<\eta<-1$. Within PHSD simulations we have found a very close value, i.e., $\mathrm{Res}(\Psi_2^{-3<\eta<-1})\simeq0.175$. 
The magnitude of the elliptic flow is correlated with the determination of the event plane and its resolution. \\
The dashed red line in Fig.~\ref{v2_ch_0-5} is the observed elliptic flow $\left\langle\cos \left[2\left(\phi-\Psi_2\right)\right]\right\rangle$ in PHSD simulations, i.e., without division for the event-plane angle resolution, showing how important is the method used to compute the elliptic flow for a meaningful comparison with experimental data.\\
The final result is comparable in magnitude with the elliptic flow found in collisions between heavy nuclei, giving indications of the fact that even in proton-nucleus collisions, despite the volume smallness and the lifetime shortness of the fireball, the formation of QGP droplets allow the generation of collective patterns visible in a preferential direction of particle emission.

\subsection{Prediction for the directed flow}
\label{v1_results}

In this section we present our prediction of the directed flow of charged and identified particles and we discuss the effect of electromagnetic fields. Indeed, the directed flow is very promising observable to investigate the influence of electromagnetic fields, which could lead to a separation of positively and negatively charged particles along the impact parameter axis.

\begin{figure}[t!]
\centering
\includegraphics[width=\columnwidth]{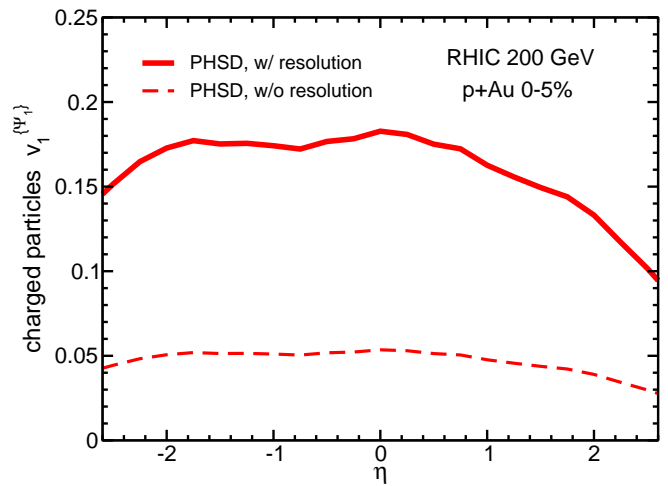}
\caption{(Color online) Pseudorapidity dependence of the directed flow of charged particles for 5\% central p+Au collisions at $\sqrt{s_{NN}}=200$ GeV obtained with PHSD simulations (red curves); the solid thick line is the $v_1^{\;\left\lbrace\Psi_1\right\rbrace}$ computed with Eq.~\eqref{vn_Psin}, whereas the dashed thin curve is the observed directed flow $\left\langle\cos \left[\phi-\Psi_1\right]\right\rangle$.}
\label{v1_ch_0-5}
\end{figure}

In Fig.~\ref{v1_ch_0-5} we plot with a solid line the $v_1^{ \; \left\lbrace \Psi_1 \right\rbrace }$ of charged particles versus pseudorapidity for 5\% central p+Au collisions at $\sqrt{s_{NN}}=200$ GeV, with the event-plane angle $\Psi_1$ computed in the pseudorapidity range $-4<\eta<-3$, where the resolution is found to be $\mathrm{Res}(\Psi_1^{-4<\eta<-3})\simeq0.293$.
We show also the observed directed flow $\left\langle\cos \left[\phi-\Psi_1\right]\right\rangle$ corresponding to the dashed line.
We found a big directed flow, with a value of about 0.18 at midrapidity; however, as for the elliptic flow, the magnitude of the directed flow as well depends on the method used for the determination of the event plane. Indeed, the $\eta$-ranges used for computing $\Psi_1$ and $\mathrm{Res}(\Psi_1)$ are regions of particle production and this ends up in correlations between $v_1$ and $\Psi_1$.

\begin{figure}[t!]
\centering
\includegraphics[width=\columnwidth]{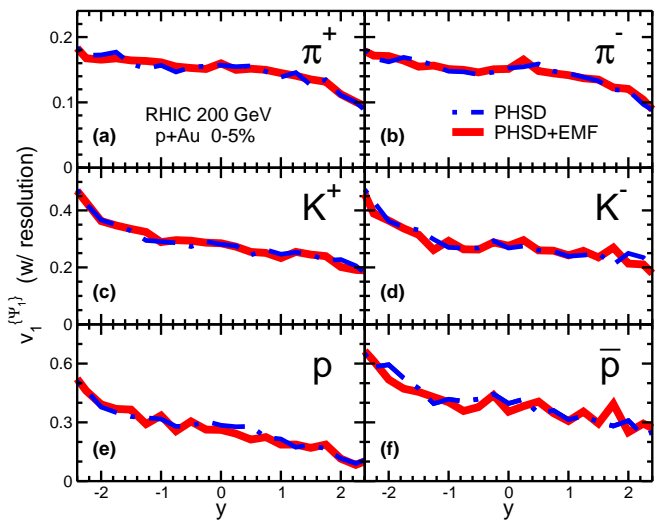}
\caption{(Color online) Directed flow of identified particles, $\pi^+$, $\pi^-$, $K^+$, $K^-$, $p$ and $\overline{p}$, as a function of rapidity for 5\% central p+Au collisions at $\sqrt{s_{NN}}=200$ GeV obtained with PHSD simulations with (solid red lines) and without (dot-dashed blue lines) electromagnetic fields.}
\label{v1_id_0-5}
\end{figure}

One main focus of this work is the investigations of the impact of electromagnetic fields and a possible effect may be a splitting of the directed flow versus rapidity of hadrons with the same mass but opposite electric charges.
In Fig.~\ref{v1_id_0-5} we plot the rapidity dependence of $v_1^{ \; \left\lbrace \Psi_1 \right\rbrace }$ of $\pi^+$, $\pi^-$, $K^+$, $K^-$, $p$ and $\overline{p}$ for 5\% central p+Au collisions at $\sqrt{s_{NN}}=200$ GeV calculated with PHSD with and without the electromagnetic fields, represented respectively by solid red lines and dot-dashed blue lines.
Within the present statistics, there is no visible difference in this observable between simulations of 5\% most central p+Au collisions with and without the electromagnetic field.\\
This could be explained in the following way. First of all we have seen in Sec. \ref{centr_select} that multiplicity fluctuations in the final state mixes events from different impact parameters and in the 0-5\% centrality bin there are contributions from peripheral to frontal collisions and events with $b\approx0$ fm get a nearly vanishing influence of electromagnetic fields, as it is clear from Fig.~\ref{fields_AuAu_pAu} (b) by looking at the values of the field components in the centre of the gold nucleus ($E_x\approx E_y\approx B_x\approx By\approx0$). Furthermore, and even more important, in experimentally measured p+Au collisions as well as in those simulated with PHSD with the ``minimum bias'' method the event plane $\Psi_1$ respect to which the directed flow is computed is very weakly correlated with the reaction plane defined by the beam axis and the impact parameter direction; this lead to a cancellation, at least partial, of opposite contributions in the directed flow.

In order to further explore this point we have performed simulations of $\sqrt{s_{NN}}=200$ GeV p+Au collisions at fixed impact parameter and computed the directed flow simply by means of Eq.~\eqref{v1} (corresponding to compute $v_1$ in each event respect to the true reaction plane). Even though the investigation of such selection would be very challenging from the experimental side, this case has the advantage to draw very clean predictions from the theoretical point of view, far from possible repercussions of the particular choices adopted for centrality selection and event-plane determination.

\begin{figure*}[t!]
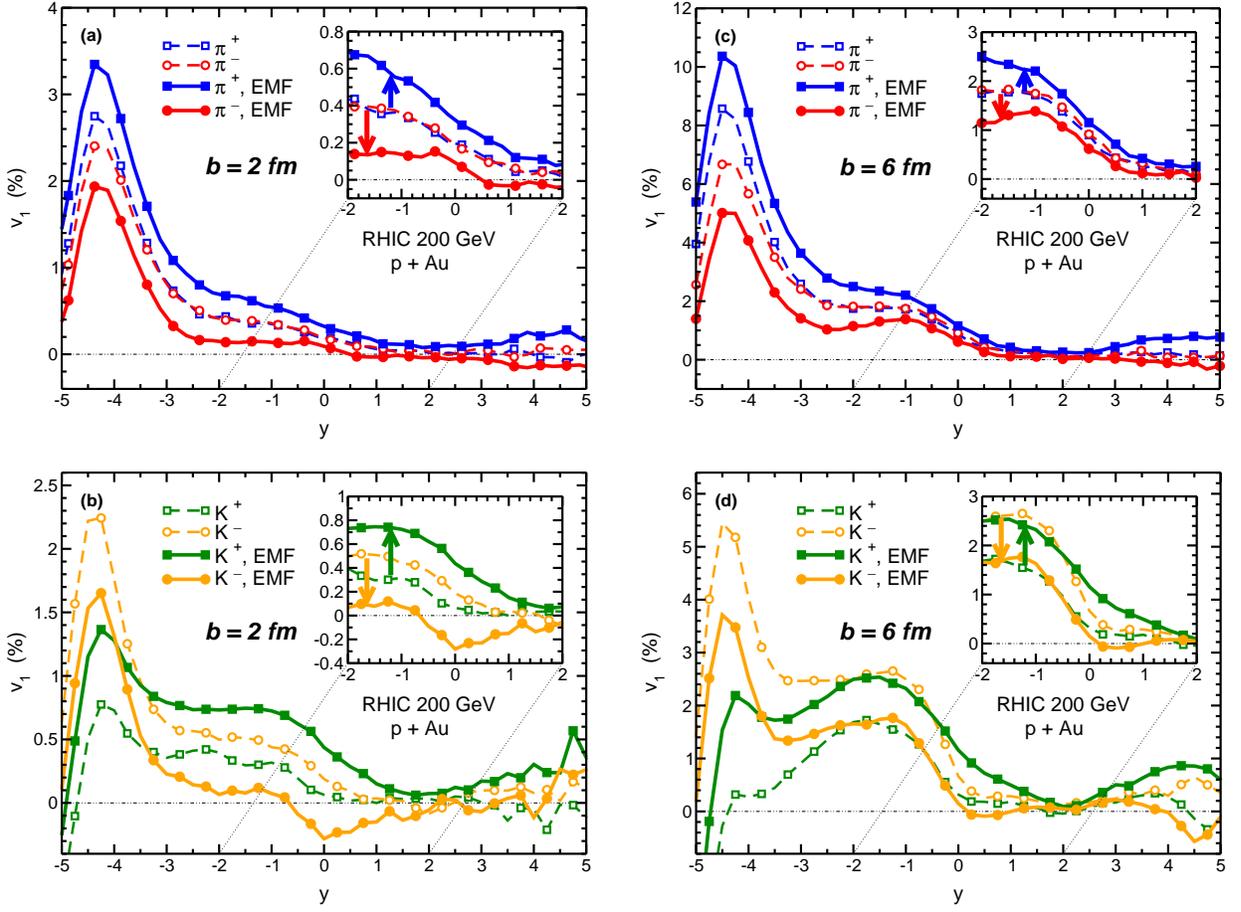

\centering
{\includegraphics[width=0.9\columnwidth]{fig_v1_y_pions_b2fm.eps}\qquad
\includegraphics[width=0.9\columnwidth]{fig_v1_y_pions_b6fm.eps}}
\\[\baselineskip]
{\includegraphics[width=0.9\columnwidth]{fig_v1_y_kaons_b2fm.eps}\qquad
\includegraphics[width=0.9\columnwidth]{fig_v1_y_kaons_b6fm.eps}}
\caption{(Color online) Directed flow of pions (panels (a) and (c)) and kaons (panels (b) and (d)) as a function of rapidity for $b=2$ fm (panels (a) and (b)) and $b=6$ fm (panels (c) and (d)) p+Au collisions at $\sqrt{s_{NN}}=200$ GeV obtained with PHSD simulations with (solid curves) and without (dashed curves) electromagnetic fields. The inset panels are zooms of the rapidity window $\left|y\right|<2$, with arrows highlighting in which direction the presence of the electromagnetic fields affect the $v_1$ observable.}
\label{v1_y_pions_kaons}
\end{figure*}

In Fig.~\ref{v1_y_pions_kaons} we plot the PHSD results for the rapidity dependence of $v_1$ (in percentage) of pions (panels (a) and (c)) and kaons (panels (b) and (d)) for p+Au collisions at $\sqrt{s_{NN}}=200$ GeV with impact parameter $b=2$ fm (panels (a) and (b)) and $b=6$ fm (panels (c) and (d)). Simulations with and without the inclusion of electromagnetic fields are labelled by solid and dashed curves respectively. Each plot shows $v_1$ in a wide rapidity window with a zoom at more central rapidities in the inset panel.
Firstly, we focus on the upper panels where the directed flow of $\pi^+$ (blue lines with squares) and $\pi^-$ (red lines with circles) is plotted. We notice that within simulations without electromagnetic fields the two oppositely-charged pions show basically the same $v_1$, with some difference only at backward rapidity $y\lesssim-3$ due to different absorption rate of $\pi^+$ and $\pi^-$ in the target region.
Switching on the electromagnetic fields in the PHSD simulations, we can clearly see that $\pi^+$ are pushed by the electromagnetic fields in the positive $x-$direction and conversely $\pi^-$ get a kick along the negative $x$, hence leading to a splitting in the $v_1$ curves of the two particles. Moreover, the push of the electromagnetic field results stronger for higher impact parameters, how can be deduced by comparing the left and right panels corresponding respectively to $b=2$ fm and $b=6$ fm.
The effect on the kaons shown in the lower panels is even more interesting because the electromagnetic fields generate a flip in the directed flow of $K^+$ (green lines with squares) and $K^-$ (orange lines with circles). Indeed, the two mesons present a different directed flow even in simulations that do not account for the electromagnetic fields. This is due to the fact that $K^+$ ($\overline{s}u$ system) receive more contributions from transported $u$ and $d$ quarks from the initial colliding nuclei with respect to $K^-$ ($s\overline{u}$ system) \cite{Dunlop:2011cf, Adamczyk:2017nxg}; hence, at backward rapidity $K^+$ displays a smaller $v_1$ with respect to $K^-$.
The electromagnetically-induced splitting in the directed flow of charged kaons is turned over respect to that discussed above and dominates over it.  

For both pions and kaons, the direction of the splitting in $v_1$ results from the contrast between sideways kicks on charged particles by electric and magnetic forces: within our convention for the reference frame, the electric part of the Lorentz force \eqref{lorentz} pushes positively charged particles along the positive $x-$direction and negatively charged particles along the negative $x-$direction while the magnetic Lorentz force does the opposite \cite{Voronyuk:2014rna, Toneev:2016bri, Gursoy:2014aka, Gursoy:2018yai, Das:2016cwd, Coci:2019nyr, Chatterjee:2018lsx}.
As highlighted by the arrows in the insets of Fig.~\ref{v1_y_pions_kaons}, the winner of this force balance in proton-nucleus reactions is the electric field, whose effects on directed flow could be distinguished as due to Faraday induction and Coulomb interaction \cite{Gursoy:2018yai}; while the Faraday effect and the Coulomb contribution within the plasma are the main origin of the $v_1$ splitting in the central rapidity window, the Coulomb force exerted by proton spectators affect mainly particle $v_1$ close to the target region, leading to an attraction of $\pi^-$ and $K^-$ and a repulsion of $\pi^+$ and $K^+$.

\begin{figure}[hbt!]
\centering
\includegraphics[width=0.98\columnwidth]{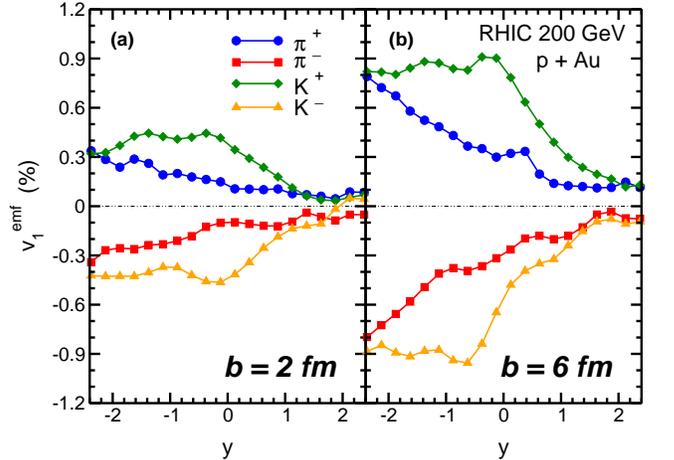}
\caption{(Color online) Directed flow induced by the electromagnetic fields for pions and kaons in p+Au collisions at $\sqrt{s_{NN}}=200$ with impact parameter $b=2$ fm (a) and $b=6$ fm (b).}
\label{v1emf_y}
\end{figure}

\begin{figure*}[t!]
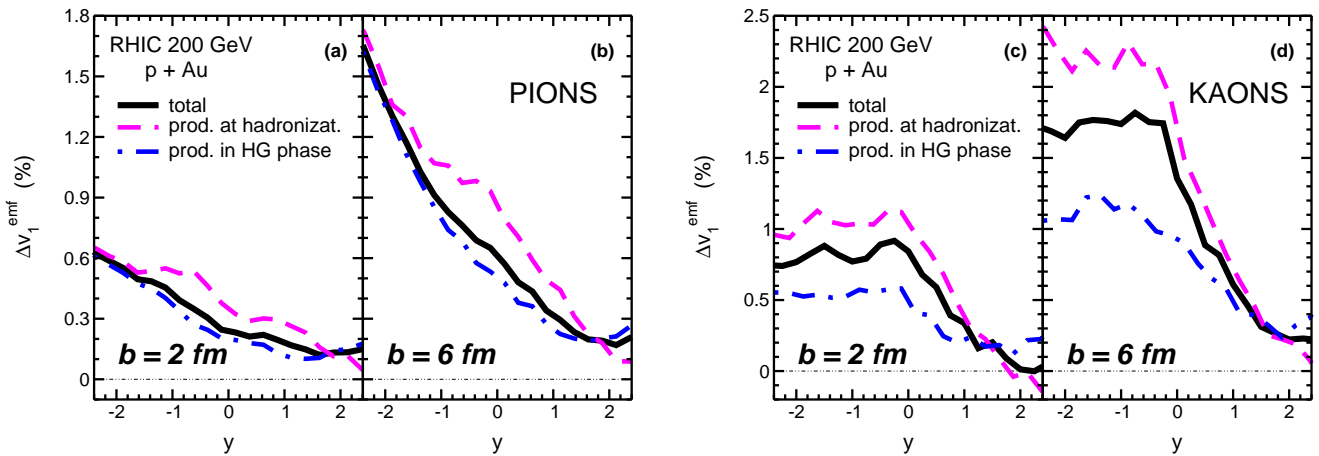

\centering
{\includegraphics[width=0.97\columnwidth]{fig_dv1emf_pions.eps}\qquad
\includegraphics[width=0.97\columnwidth]{fig_dv1emf_kaons.eps}}
\caption{(Color online) Splitting of directed flow between positively and negatively charged mesons induced by the electromagnetic fields estimated by Eq.~\eqref{dv1_emf} for pions (panels (a) and (b)) and kaons (panels (c) and (d)) in p+Au collisions at $\sqrt{s_{NN}}=200$ with impact parameter $b=2$ (panels (a) and (c)) fm and $b=6$ fm (panels (b) and (d)). In each panel the total splitting (solid black lines) is shown along with the contributions given by mesons coming directly from hadronization of the QGP (dashed magenta curves) and mesons produced in the hadron gas (HG) phase (dot-dashed blue curves).}
\label{dv1_y}
\end{figure*}

In order to pinpoint the magnitude of the directed flow of $\pi^+$, $\pi^-$, $K^+$, $K^-$ induced by the electromagnetic field in p+Au collisions at $\sqrt{s_{NN}}=200$, we show in Fig.~\ref{v1emf_y} (in percentage) for each particle species the quantity
\begin{equation}
v_1^{emf}\equiv v_1^{(PHSD+EMF)}-v_1^{(PHSD)}
\end{equation}
i.e., the difference of $v_1$ in PHSD simulations with and without the electromagnetic field, therefore removing the directed flow due to other causes such as vortical effects and fluctuations. We notice that the effect increases for increasing impact parameter, as it is evident comparing the simulations at $b=2$ fm (a) to those at $b=6$ fm (b), and it is bigger in magnitude for kaons (green diamonds and orange triangles) respect to pions (blue circles and red lines squares), at least in the rapidity region $\left|y\right|<2$. This could be attributed to the different mass and hence different velocity of the two meson species: indeed, the faster the particle, the stronger is the force exerted on it by the magnetic field and this in turn leads to an higher compensation between electric and magnetic pushes and then to a smaller total effect on $v_1$; moreover, for slower particles is bigger the duration of the influence of the electromagnetic fields.

The magnitude of the splitting in the directed flow of hadrons with opposite charge can be measured by the quantity $\Delta v_1\equiv v_1^+-v_1^-$, where $v_1^+$ and $v_1^-$ are the directed flow of the positively and negatively charged particles respectively.
Then we consider the quantity
\begin{equation}\label{dv1_emf}
\Delta v_1^{emf}\equiv\Delta v_1^{(PHSD+EMF)}-\Delta v_1^{(PHSD)}
\end{equation}
which gives information on the magnitude of the directed flow splitting induced by the electromagnetic fields.
This quantity (in percentage) is presented in Fig.~\ref{dv1_y} for p+Au collisions at $\sqrt{s_{NN}}=200$ with two different impact parameter $b=2$ fm (panels (a) and (c)) and $b=6$ fm (panels (b) and (d)); the results for pions and kaons are shown respectively in the left and right panels by solid black curves.\\
In order to understand how much of this electromagnetically induced splitting keeps trace of the splitting produced at partonic level, we have distinguished between mesons created directly through hadronization of quark-gluon plasma (dashed magenta lines) and mesons produced in the hadronic phase (dot-dashed blue lines); see channel decomposition in Fig.~\ref{dNdy_channels}.
We see that in the rapidity window $\left|y\right|<2$, for both pions and kaons, the splitting generated by the electromagnetic fields at partonic level is higher than that induced in the hadronic phase; this difference is far more pronounced in the strange sector, where the splitting induced in the QGP dominates over that produced in the confined phase for about a factor of two at backward rapidity.\\ In all cases the electromagnetic signal becomes weaker going in the forward direction, due to the fact that from one hand particle production is smaller in this region and from the other hand all spectators come from the Au nucleus, hence the fields produced by participants and spectators imprints their influence mainly at backward rapidity.
Furthermore, we notice that the electromagnetically induced splitting increases with increasing impact parameter following the increasing trend of $E_x$.

\section{Conclusions}
\label{conclusions}

In this work we have studied p+Au collisions at $\sqrt{s_{NN}}=200$ GeV with the PHSD approach, which describes the entire dynamical evolution of the collision and includes in a consistent way the dynamical generation of retarded electromagnetic fields and their influence on quasi-particle propagation.

We have analysed the electromagnetic field generated by all charged particles, both spectator protons and charged hadronic and partonic particles produced in the collision.
We have shown the distribution in the transverse plane of the transverse components of the field as well as their time evolution. 
The $x-$component of the electric field is comparable in magnitude to $B_y$, that in symmetric colliding systems is the only dominant component of the electromagnetic field.
Both $E_x$ and $B_y $ are strongly asymmetric inside the overlap region and decrease very fast, approaching zero after about 0.25 fm/$c$ from the first nucleon-nucleon collision.

We have performed simulations by means of PHSD with and without the inclusion of electromagnetic fields and compared the corresponding outcomes in order to disentangle the possible impact of the fields on final observables.

The PHSD results for the charged particle rapidity distribution fairly agree with the experimental data from the PHENIX Collaboration: particle production is enhanced in the Au-going directions and the asymmetry between forward and backward rapidity increases with the centrality of the collision.
Moreover, we have shown our predictions for rapidity densities and momentum spectra of pions, kaons, protons and antiprotons, highlighting also that these observables are not modified by the presence of electromagnetic fields.

We have studied the first two flow harmonics of the azimuthal particle distribution of charged and identified particles.
We have found a good agreement of the elliptic flow of charged particle in the 5\% most central collisions computed with PHSD with respect to the PHENIX experimental result.
We have shown the sensitivity of the result on the event-plane reconstruction accounting for the resolution in line with the experimental procedure. \\ 
Furthermore, we have presented our prediction of the directed flow $v_1$ of charged and identified particles for collisions in the 0-5\% centrality bin as well as for collisions at fixed impact parameter $b=2$ fm and $b=6$ fm.
Distributions of particles with the same mass but opposite electric charge could be splitted by the electromagnetic fields and we have clearly observed this effect in the directed flow of pions and kaons in collisions at fixed impact parameter: the $v_1$ of $\pi^+$ and $K^+$ is pushed upward and the $v_1$ of $\pi^-$ and $K^-$ is pushed downward with respect to the case without electromagnetic fields. This trend is visible in a wide rapidity window, but is more pronounced in the Au-going side and the splitting increases for more peripheral collisions.

Moreover, we have investigated the amount of splitting generated in the partonic and hadronic stages, distinguishing between mesons formed by hadronization of the quark-gluon plasma or produced by hadronic interaction. At rapidities $\left|y\right|<2$ the $v_1$ generated at partonic level is higher than that built up in the confined phase; for the strange mesons the first contribution is rather dominant over the latter at backward rapidity (Au-going side).

Thus, we conclude that the study of the directed flow of charged hadrons can shed light on the influence of electromagnetic fields on the dynamics of proton-induced collisions.

\section*{Acknowledgements}
The authors appreciate useful discussions with Wolfgang Cassing, Darren McGlinchey, Ilya Selyuzhenkov, Olga Soloveva, Taesoo Song and Qiao Xu.
L.O. and E.B. acknowledge support by the Deutsche Forschungsgemeinschaft (DFG) through the grant CRC-TR 211 'Strong-interaction matter under extreme conditions', from the COST Action THOR CA15213 and by the Deutscher Akademischer Austauschdienst (DAAD).
L.O. has been in part financially supported by the Alexander von Humboldt Foundation.
The computational resources have been provided by the LOEWE-Center for Scientific Computing and the Green IT Cube at GSI.

\bigskip

\pagebreak[4]


%
%
%
%


\bibliography{References}

\end{document}